\documentclass[12pt]{article}

\usepackage{latexsym}

\begin{document}
%%%%%%%%%%%%%%%%%%%%%%%%%%%%%%%%%%%%%%%%%%%%%%%%%%%%%%%%%%%%%%%%%%%

\title{Distorted charged dilaton black holes}

\author{
     Stoytcho S. Yazadjiev \thanks{E-mail: yazad@phys.uni-sofia.bg}\\
{\footnotesize  Department of Theoretical Physics,
                Faculty of Physics, Sofia University,}\\
{\footnotesize  5 James Bourchier Boulevard, Sofia~1164, Bulgaria }\\
}

\date{}

\maketitle

\begin{abstract}
We construct exact static, axisymmetric solutions of
Einstein-Maxwell-dilaton gravity presenting distorted charged
dilaton black holes. The thermodynamics of such distorted black
holes is also discussed.
\end{abstract}

%%%%%%%%%%%%%%%%%%%%%%%%%%%%%%%%%%%%%%%%%%%%%%%%%%%%%%%%%%%%%%%%%%%

%\draft
\sloppy
%\scrollmode
%%%%%%%%%%%%%%%%%%%
\renewcommand{\baselinestretch}{1.3} %
\newcommand{\sla}[1]{{\hspace{1pt}/\!\!\!\hspace{-.5pt}#1\,\,\,}\!\!}
\newcommand{\db}{\,\,{\bar {}\!\!d}\!\,\hspace{0.5pt}}
\newcommand{\partb}{\,\,{\bar {}\!\!\!\partial}\!\,\hspace{0.5pt}}
\newcommand{\dsla}{\partb}
\newcommand{\eql}{e _{q \leftarrow x}}
\newcommand{\eqr}{e _{q \rightarrow x}}
\newcommand{\ite}{\int^{t}_{t_1}}
\newcommand{\itz}{\int^{t_2}_{t_1}}
\newcommand{\itd}{\int^{t_2}_{t}}
\newcommand{\lfrac}[2]{{#1}/{#2}}
\newcommand{\dV}{d^4V\!\!ol}
\newcommand{\ben}{\begin{eqnarray}}
\newcommand{\een}{\end{eqnarray}}
\newcommand{\la}{\label}

%%%%%%%%%%%%%%%%%%%
\section{Introduction}

The notion of distorted black holes is very natural from physical
point of view. The distorted  black hole can be viewed as an
isolated system with a black hole in the center and a matter
distribution at a finite distance from the black hole. The
surrounding matter influences the black hole and thus the latter
will be distorted.

Distorted black holes within the framework of general relativity
have been considered by many authors (see, for example, \cite{GH}
- \cite{X}). Geroch and Hartle \cite{GH} have shown how to
construct general relativistic solutions presenting distorted
black holes and have established their global structure,
thermodynamic behavior and their evolution with the emission of
Hawking radiation. It turns out that the spherical and toroidal
black holes are the only possibilities for the topology of
horizon cross sections \cite{GH}.

Very recently, charged distorted black holes in Einstein-Maxwell
gravity have been discussed in \cite{FK}. The authors have
presented new static, axisymmetric solutions to Einstein-Maxwell
gravity generalizing the uncharged solutions studied in \cite{GH}.
The zeroth and the first laws of charged distorted black holes
have also been discussed.

The aim of the present work is to construct exact solutions
presenting distorted charged dilaton black holes within the
framework of a low energy string model - the so called
Einstein-Maxwell-Dilaton (EMD) gravity. The thermodynamics of
distorted dilaton black holes will also be discussed.

In the case of vacuum, static and axisymmetric space-times,
Einstein equations reduce in practice to the Laplace equation on
flat space. Making use of the linearity of the Laplace equation,
distorted black holes solutions can be obtained in a simple way
by adding an appropriate distortion function to the solution
presenting an isolated Schwarzschild black hole. This method,
however, can't be applied to the EMD gravity because even for
static,  axisymmetric space-times the EMD equations are a highly
nonlinear set of coupled partial differential equations.
Fortunately, after a dimensional reduction, the static EMD
equations possess a large group of an internal symmetry which can
be employed to construct exact solutions \cite{Y1}, \cite{Y2}.
The group action transforms the distorted Schwarzschild solutions
into new static, axisymmetric solutions of EMD equations. These
new solutions present distorted charged dilaton black holes and
depend on three different parameters. One of the parameters comes
directly from the distorted Schwarzschild solution, while the
remaining are group parameters. From a physical point of view the
free parameters describe how different kinds of external matter
affect the black hole.

\section{Static axisymmetric EMD equations}

In the so called Einstein frame the EMD gravity is described by
the following system of equations:

\ben \la{FE} &R_{ab}& = 2\partial_{a}\varphi\partial_{b}\varphi +
2e^{-2\varphi}\left(F_{ac}F_{b}{}^{c} -
{1\over4}g_{ab}F_{cd}F^{cd}\right)
\\ \nonumber &\Box\varphi & = - {1\over 2}e^{-2\varphi}F_{ab}F^{ab}
\\ \nonumber &\nabla_{b}&\left(e^{-2\varphi}F^{ab}\right) = 0 \, .\een

Here $R_{ab}$ is the Ricci tensor with respect to the space time
metric $g_{ab}$, $F_{ab}=(dA)_{ab}$ is the Maxwell 2-form and
$\varphi$ is the dilaton field.

The metric of a static and axisymmetric space-time has the Weyl
form

\ben ds^2 = -e^{2u}dt^2 + e^{2(h - u)}\left(d\rho^2 + dz^2\right)
+ e^{-2u}\rho^2d\phi^2 \een

where $u$ and $h$ are functions of $\rho$ and $z$ only. The metric
admits two Killing vectors $\xi = {\partial \over \partial t }$
and $\eta = {\partial \over \partial\phi}$.

In what follows we consider only the electrically charged case.
Then the Maxwell 2-form can be cast in the form

\ben F =  e^{-2u} d\Phi\wedge \xi  .\een

Here we have denoted the electric potential by $\Phi$.

Pure magnetic solutions  can be obtained by performing the
discrete duality transformation

\ben e^{-2\varphi}F \to \star F, \; \varphi \to - \varphi, \;
g_{ab}\to g_{ab} \een

which is a symmetry of (\ref{FE}).

In the static, axisymmetric case the EMD gravity equations are
expressed in terms of the metric functions $u$ and $h$, the
electric potential $\Phi$ and the dilaton $\varphi$. The
resulting equations are:

\ben \la{SASEMD}
\partial^2_{\rho}u + {1\over \rho}\partial_{\rho}u +
\partial^2_{z}u = e^{-2u -2\varphi}\left((\partial_{\rho}\Phi)^2 +
(\partial_{z}\Phi)^2 \right) \nonumber \\
\partial^2_{\rho}\varphi + {1\over \rho}\partial_{\rho}\varphi +
\partial^2_{z}\varphi = e^{-2u -2\varphi}\left((\partial_{\rho}\Phi)^2 +
(\partial_{z}\Phi)^2 \right) \nonumber \\
\partial_{\rho}\left(\rho \partial_{\rho}\Phi e^{-2u - 2\varphi}\right)
+ \partial_{z}\left(\rho \partial_{z}\Phi e^{-2u -
2\varphi}\right)=0 \\ {1\over \rho}\partial_{\rho}h=
(\partial_{\rho}u)^2 - (\partial_{z}u)^2 +
(\partial_{\rho}\varphi)^2 - (\partial_{z}\varphi)^2 - e^{-2u
-2\varphi}\left((\partial_{\rho}\Phi)^2 - (\partial_{z}\Phi)^2
\right) \nonumber \\ {1\over \rho}\partial_{z}h = 2
\left(\partial_{\rho}u\partial_{z}u +
\partial_{\rho}\varphi\partial_{z}\varphi - e^{-2u -
2\varphi}\partial_{\rho}\Phi\partial_{z}\Phi \right) . \nonumber
 \een

As it has been shown in \cite{Y1}, \cite{Y2} the static EMD
equation have $GL(2,R)$ as a group of a global internal symmetry.
This symmetry becomes transparent if we introduce the following
symmetric matrix:

\ben \la{MAT} S = \pmatrix{e^{2u} -2\Phi^2 e^{-2\varphi} &
-\sqrt{2}\Phi e^{-2\varphi}\cr -\sqrt{2}\Phi e^{-2\varphi} & -
e^{-2\varphi}} . \een

Because the group $GL(2,R)$ is not connected we may restrict
ourselves to the one of the connected components, say
$GL^{(+)}(2,R)$.

In terms of the matrix $S$ the static, axisymmetric EMD equations
are written in the following compact form:

\ben  \partial_{\rho}\left(\rho S^{-1}\partial_{\rho} S\right) +
\partial_{z}\left(\rho S^{-1}\partial_{z}S \right) = 0 \nonumber
\\ {1\over \rho}\partial_{\rho}h = {1\over 4}Tr\left(\partial_{z}S^{-1}\partial_{z}S -
 \partial_{\rho}S^{-1}\partial_{\rho}S\right) \\
 {1\over \rho}\partial_{z}h = - {1\over
 2}Tr\left(\partial_{\rho}S\partial_{z}S^{-1}\right).
 \nonumber
\een

The equation for the matrix $S$ is just the chiral equation. It
is, therefore, obvious that the static, axisymmetric EMD equations
form a completely integrable system. There are powerful soliton
methods for solving the chiral equation. In the present work,
however we shall not solve this equation directly. Instead, we
shall apply the symmetry group to construct exact solutions
presenting distorted charged dilaton black holes from uncharged
general relativistic ones.

First, we shall demonstrate how the solution describing an
isolated charged dilaton black hole can be generated from a
Schwarzschild one.

The group $GL(2,R)$ acts as follows:

\ben S \to ASA^{T} \een

where $A \in GL(2,R)$.

The action of the symmetry group does not, in general, preserve
the asymptotic flatness. That is why when we are concerned with
isolated black holes, we will need the $GL^{(+)}(2,R)$-subgroup
preserving asymptotic flatness. The desired subgroup can be easily
found. In order for $S$ to present an asymptotically flat
spacetime it is necessary\footnote{Without loss of generality we
take $\varphi(\infty)=0$. } that $S(\infty) = \sigma_{3}$ at
spatial infinity. Here $\sigma_{3}$ is the third Pauli matrix. The
elements of the subgroup should then satisfy $$B\sigma_{3}B^{T}=
\sigma_{3}.$$ Therefore, the subgroup preserving the asymptotic
flatness is $SO(1,1)$.

The parameterization of $SO(1,1)$ is taken to be the standard one:

 \ben B = \pmatrix{ {1\over \sqrt{1 - \beta^2}}  & {\beta\over
\sqrt{1 - \beta^2}}   \cr {\beta\over \sqrt{1 - \beta^2}}  &
{1\over \sqrt{1 - \beta^2}}}   \; . \een

In order to generate the charged dilaton black hole solution we
start with the Schwarzschild solution written in the standard
spherical coordinates,

\ben ds^2 = - (1 - {2M\over r})dt^2 + {dr^2 \over 1 - {2M\over r}}
+ r^2d\Omega^2\,\,\,  \een

where $d\Omega^2= d\theta^2 + \sin^2(\theta)d\phi^2$. The
relationship between the coordinates $r$ and $\theta$ and the
Weyl coordinates is given  by equations (\ref{SW1}) - (\ref{SW3})
below.

The matrix $S$ which corresponds to the Scwarzschild solution is

\ben S_{S}= \pmatrix{e^{2\lambda_{S}} & 0 \cr 0 & -1}\een

where $e^{2\lambda_{S}}= 1 - {2M\over r}$ .

The desired solution is presented by the matrix

\ben S = BS_{S}B^{T} = {1\over 1 -
\beta^2}\pmatrix{e^{2\lambda_{S}} - \beta^2 &
\beta\left(e^{2\lambda_{S}} - 1\right)    \cr
\beta\left(e^{2\lambda_{S}} - 1\right) & \beta^2 e^{2\lambda_{S}}
- 1} . \een

Recovering the four-metric we have

\ben ds^2= - {1 - {2M\over r} \over 1 + {2\beta^2M\over (1 -
\beta^2)r }}dt^2 + e^{-2\varphi}\left({dr^2 \over 1 - {2M\over r}
}  + r^2 d\Omega^2 \right) \een

where \ben e^{-2\varphi}= 1 + {2\beta^2M \over (1-\beta^2)r} \;
.\een

The electric potential is, respectively, \ben \Phi =
{\sqrt{2}\beta M \over 1 -\beta^2} {1\over r + {2\beta^2M\over
1-\beta^2}} \; .\een

The obtained solution describes an isolated charged dilaton black
hole with mass ${\cal M} = {M\over 1 - \beta^2}$ and charge $Q =
{\sqrt{2}\beta \over 1 - \beta^2}M$. It may be cast in more
familiar form by performing the coordinate shift $R = r +
{Q^2\over {\cal M}}$. Then we obtain the black hole solution in
Garfinkle-Horowitz-Strominger coordinates:

 \ben ds^2 &=& - (1 - {2{\cal M}\over R })dt^2 + (1 -
{2{\cal M}\over R })^{-1} d\,R^2 + R\,(R - {Q^2 \over {\cal M}})
d\Omega^2 \nonumber \\ e^{2\varphi} &=& 1 - {Q^2 \over {\cal M}R}
\\
 \Phi &=& {Q\over R } \nonumber .\een

We shall follow a similar line of thoughts in the next section to
construct distorted charged dilaton black hole solutions from
general relativistic ones.

The matrix $B$ transforming the Schwarzschild solution into the
solution presenting charged dilaton black holes with mass ${\cal
M}$ and charge $Q$ can be written as

\ben B_{(M,Q)} = {1\over \sqrt{1 - {Q^2 \over 2{\cal M}^2}}}
\pmatrix{1 & {Q\over \sqrt{2}{\cal M}} \cr  {Q\over \sqrt{2}{\cal
M}} & 1 }\; . \een

\section{Distorted black holes}

This section is devoted to the construction of charged dilaton
solutions presenting distorted black holes. For the reader's
convenience we begin with a brief review of the uncharged general
relativistic distorted black holes following \cite{FK} closely.
Finally, some interesting properties of the distorted charged
black holes will be discussed.

\subsection{Distorted Schwarzschild black hole}

The vacuum, static and axisymmetric Einstein equations reduce to
the following system of partial differential equations for the
metric functions \footnote{In this subsection we write $\lambda$
instead of u.}  $\lambda$ and $h$:

\ben \la{VSAEE}
\partial^2_{\rho}\lambda + {1\over \rho}\partial_{\rho}\lambda +
\partial^2_{z}\lambda = 0 \nonumber \\
 {1\over \rho}\partial_{\rho}h=
(\partial_{\rho}\lambda)^2 - (\partial_{z}\lambda)^2
\\ {1\over \rho}\partial_{z}h = 2\partial_{\rho}\lambda\partial_{z}\lambda
\; .\nonumber  \een

If a solution for $\lambda$ has been found then the second metric
function $h$ is obtained by integrating the remaining equations.
The function $h$ is determined up to a constant. The constant can
be fixed by imposing the boundary condition:  $h=0$ on the $\rho =
0$ axis at all points where $\lambda$ is regular. This condition
also ensures that, near the axis, the orbits of the Killing vector
$\eta$ are closed curves of period $2\pi$.

The exterior Schwarzschild solution is a vacuum Weyl solution. The
explicit form of the Schwarzschild solution presenting black hole
with mass $M$ is

\ben \la{SW1} 2\lambda_{S} = \ln\left({L - M\over L + M}
\right)\,\, and \,\,2h_{S}= \ln\left( {L^2 - M^2 \over L^2 -
\Delta^2} \right) \een

where $L$ and $\Delta$ are functions of $\rho$ and $z$ given by

\ben \la{SW2}  L = {1\over 2}\left(\sqrt{\rho^2 + (z + M)^2} +
\sqrt{\rho^2 + (z - M)^2} \right),  \nonumber \\ \Delta= {1\over
2}\left(\sqrt{\rho^2 + (z + M)^2} - \sqrt{\rho^2 + (z - M)^2}
\right).\een

In Weyl coordinates, the horizon ${\cal H}$ is placed in the
segment $\mid z \mid \le M$ on the axis $\rho = 0$. As it can be
seen the metric functions $\lambda_{S}$ and $h_{S}$ are well
behaved everywhere except in the limit $\rho \to 0$ for the
segment $\mid z \mid \le M$ where they diverge logarithmically.
There is, however, nothing dangerous because the divergence is
just the well-known coordinate singularity. This can be seen by
performing the following coordinate transformation:

\ben \la{SW3} r = L + M, \,\, z = L \cos(\theta) \nonumber\\
\cos(\theta)= {\Delta \over M}, \,\,\; \rho^2 = (L^2 -
M^2)\sin^2(\theta) \een

Under this transformation the solution takes the standard
Schwarzschild form with an event horizon ${\cal H}$ placed at
$r=2M$.

Distorted black holes solutions are constructed as follows. The
vacuum, static and axisymmetric  Einstein equations (\ref{VSAEE})
are linear in the function $\lambda$. This fact allows us to
construct new solutions simply by adding any harmonic potential
$\lambda_{D}$ to the Schwarzschild potential $\lambda_{S}$. The
solutions obtained in this way  can be viewed as distorted analogs
of the Schwarzschild solution. When the potential $\lambda_{D}$
is everywhere regular the location of the horizon ${\cal H}$ will
be unchanged. It should be noted, however, that since
$\lambda_{D}$ is a harmonic function and is regular on the
horizon it cannot tend to zero at infinity. Moreover,
$\lambda_{D}$ must diverge at infinity and therefore the solution
will not be asymptotically flat.

Having the new potential $\lambda = \lambda_{S} + \lambda_{D}$,
the other metric function $h$ is obtained by integrating the
remaining equations of (\ref{VSAEE}) with the boundary condition
that has been imposed. It turns out that the boundary condition
places a restriction on the distorting potential $\lambda_{D}$,
too. In order for $h$ to vanish on both disconnected sections of
the axis, it is required that $\lambda_{D}$ takes the same value
$u_{*}$ at both ends of the segment ${\cal H}$. Moreover, it turns
out that

\ben h\mid_{\cal H} = 2\lambda_{D}\mid_{\cal H} - 2u_{*} .\een

This important relation allows us to show that the metric is
regular on the horizon.

Finally, the distorted metric can be presented in Schwarzschild
coordinates as follows: \ben ds^2 \!=\! -
e^{2\lambda_{D}}\!\left(1 - {2M\over r} \right)dt^2  + e^{2(h_{D}-
\lambda_{D})}\!\left({dr^2\over 1 - {2M\over r}} +
r^2d\theta^2\right)+ e^{-2\lambda_{D}}r^2\sin^2(\theta)d\phi^2
\een

where we have set $h_{D}= h - h_{S}$.

As has been shown in \cite{GH}, this metric can be analytically
continued through the horizon.

As we have already mentioned the distorted solutions are not
asymptotically flat provided the function $\lambda_{D}$ is
everywhere harmonic and is not a constant. As shown in \cite{GH}
it is possible to find asymptotically flat extensions if we
require that $\lambda_{D}$ is harmonic only in a neighborhood of
the horizon and to extend $\lambda_{D}$ and $h_{D}$ so that they
vanish at infinity. Therefore, we assume that in the intervening
region the vacuum Einstein equations are not satisfied, i.e.
there is some kind of matter present in this region. The
distortion of the black hole is caused by this matter. It was
shown that if the matter satisfies the strong energy condition
the function $\lambda_{D}$ must be non-positive everywhere in
space-time and therefore $u_{*}\le 0$ (see \cite{GH}).

For more details concerning distorted Schwarzschild black holes we
refer the reader to \cite{GH} and \cite{FK}.

\subsection{Distorted charged dilaton black holes}

The existence of a group of an internal symmetry for the
dimensionally reduced EMD equations allows us to transform any
given static solution of vacuum Einstein equations to a static
solution of EMD equations. Therefore, the vacuum Weyl solutions
describing distorted Schwarzschild black holes can be transformed
into Weyl EMD solutions. These new solutions, as it is natural to
expect, should present distorted charged dilaton black holes.

The group action transforming a distorted Schwarzschild solution
into a static, axisymmetric solution of EMD equations involve
arbitrary $GL^{(+)}(2,R)$ matrices and we need to specify them.
The $GL^{(+)}(2,R)$ matrices should be of the form

\ben \la{MG} A = \pmatrix{e^{\upsilon_{1}} & 0 \cr 0 &
e^{\upsilon_{2}} } B_{({\cal M}, Q)} \een

where $\upsilon_{1}$ and  $\upsilon_{2}$ are arbitrary constants.
This is the most general form of the $GL^{(+)}(2,R)$ matrices
which are physically interesting in the present context because
the remaining transformation, which is not included in (\ref{MG}),
is a pure electromagnetic gauge.

Starting with a distorted Schwarzschild solution presented by the
matrix

\ben S_{DS} = \pmatrix{e^{2\lambda} & 0 \cr 0 & -1} \een

where $\lambda = \lambda_{S} + \lambda_{D}$, we obtain a new
static, axisymmetric EMD solution given by the matrix

\ben S= AS_{DS}A^{T}\,\,\, .\een

It should be noted that  the new EMD solution has the same metric
function $h$ as the original distorted Schwarzschild solution ($h
= h_{DS}$).

The space-time metric can be recovered and its form is

\ben \la{DEMD}ds^2 = - e^{2u_{D}}\left(1 - {2{\cal M}\over R}
\right)dt^2 + e^{2(h_{D} - u_{D})} \left[{dR^2 \over 1 - {2{\cal
M}\over R}}
 + R\left(R - {Q^2 \over {\cal M}}\right)d\theta^2\right] +
 \nonumber \\
  e^{-2u_{D}}R \left(R - {Q^2 \over {\cal M}} \right)\sin^2(\theta)d\phi^2 \een

where the distorting potential $u_{D}$ is given by

\ben \la{GP} e^{2u_{D}} = e^{2\upsilon_{1}}{ \left(1 - {Q^2 \over
2 {\cal M}^2}e^{2\lambda_{S}}\right) e^{2\lambda_{D}} \over 1 -
{Q^2 \over 2 {\cal M}^2}e^{2\lambda_{S}}e^{2\lambda_{D}} }\, .
\een

To specify the solution completely we should also give the dilaton
 and electromagnetic fields. They are given, respectively, by the
 following expressions:

\ben \la{DP} e^{2\varphi}= \left(1- {Q^2\over {\cal M}R} \right)
e^{2\varphi_{D}} = \left(1- {Q^2\over {\cal M}R}
\right)e^{-2\upsilon_{2}}     { \left(1 - {Q^2 \over 2 {\cal
M}^2}e^{2\lambda_{S}}\right) \over 1 - {Q^2 \over 2 {\cal
M}^2}e^{2\lambda_{S}}e^{2\lambda_{D}} }\, \, \, ,\een

\ben \la{EP} \Phi  = {Q\over 2{\cal M}}e^{\upsilon_{1} -
\upsilon_{2}} { 1 - e^{2\lambda_{S}}e^{2\lambda_{D}} \over 1 -
{Q^2\over 2{\cal M}}e^{2\lambda_{S}}e^{2\lambda_{D}} }\,\,\, .\een

It is useful to express the electric potential $\Phi$ as the
standard potential ${Q\over R}$ plus a distortion part $\Phi_{D}$:

\ben \Phi = {Q\over R } + \Phi_{D} = {Q\over R } + {Q\over 2{\cal
M}}\left[e^{\upsilon_{1}- \upsilon_{2}} {1 -
e^{2\lambda_{S}}e^{2\lambda_{D}}\over 1 - {Q^2\over 2{\cal
M}}e^{2\lambda_{S}}e^{2\lambda_{D}} } -  {1 -
e^{2\lambda_{S}}\over 1 - {Q^2\over 2{\cal M}}e^{2\lambda_{S}} }
\right] \, .
 \een

It is easy to see that one has

\ben u_{D}\mid_{\, {\cal H}}\, = \lambda_{D}\mid_{\, {\cal H}} +
 \; \upsilon_{1}. \een

on the horizon.

Therefore, we obtain that the following is satisfied on the
horizon:

\ben \left( h_{D} - 2u_{D}\right)\mid_{\, {\cal H}}\,= -2(u_{*} +
\upsilon_{1}). \een

Since $\lambda_{D}$ is a regular harmonic function the metric
(\ref{DEMD}) is non-singular in a neighbourhood of the horizon. In
the case when $\lambda_{D}$ is non-positive the metric is
non-singular everywhere outside the horizon.

It remains to show that the metric (\ref{DEMD}) is regular on the
horizon. This can be demonstrated by standard techniques. For this
purpose we introduce the new Edington-Finkelstein-like coordinate
$\omega$, given by  $d\omega = dt + e^{-2(u_{*} +\upsilon_{1}
)}{dR \over 1 - {2{\cal M }\over R}}$. In terms of the coordinate
$\omega$  the metric (\ref{DEMD}) is written in the form

\ben ds^2 &= - e^{2u_{D}}\left(1 - {2{\cal M}\over R}
\right)d\omega^{2} + 2e^{-2(u_{*} + \upsilon_{1})}e^{2u_{D}}dR
d\omega  + e^{2u_{D}} \left( e^{2(h_{D} -2u_{D})} - e^{-4(u_{*} +
\upsilon_{1})}\over 1 - {2{\cal M}\over R } \right)dR^2 \nonumber
\\ & + R\left(R - {Q^2\over {\cal M}} \right)\left[e^{2(h_{D} -
u_{D})}d\theta^2 + e^{-2u_{D}}\sin^2(\theta)d\phi^2 \right]
 \een

It is not difficult to show that the coefficient of the $dR^2$
term remains finite on the horizon. Therefore, we can conclude
that the metric (\ref{DEMD}) is regular on the horizon.

As it is clear our space-time is not asymptotically flat. Our
solution, however, can be extended {as in \cite{GH}} to be
asymptotically flat. This can be done as follows. We assume that
the solution is valid only in a neighbourhood of the horizon.
Outside this neighbourhood the solution can be extended arbitrary
to present asymptotically flat space-time. In the intervening
region the EMD equations will not be satisfied, which shows the
presence of external matter causing the black hole distortion.

\subsection{Properties of distorted charged dilaton black holes}

The horizon geometry is presented by the metric

\ben ds_{{\cal H}}^2 = R_{{\cal H}}^2\left(1 - {Q^2\over {\cal
M}R_{{\cal H}}} \right) \left[ e^{2h_{D}(\theta) - 2u_{D}(\theta)
} d\theta^2 + e^{ -
2u_{D}(\theta)}\sin^2(\theta)d\phi^2\right]\een

where $h_{D}(\theta)=h_{D}(R_{\cal H}, \theta)$, $u_{D}(\theta) =
u_{D}(R_{\cal H}, \theta)$ and $R_{\cal H}= 2{\cal M}$. This is an
axisymetric but not in general spherically symmetric metric on a
topological 2-sphere. Therefore, the geometry of the horizon is
distorted. In order to see this in more explicit form we write the
curvature of the horizon 2-sphere

\ben {}^{2}R = 2e^{4(u_{*} + \upsilon_{1})}  {e^{-2u_{D}(\theta)}
\over R_{{\cal H}}^2\left(1 - {Q^2\over {\cal M}R_{{\cal H}}}
\right) } \left( 1 - 2 (\partial_{\theta}u_{D}(\theta))^2 +
3\cot(\theta)\partial_{\theta}u_{D}(\theta) + \partial^2_{\theta}
u_{D}(\theta)\right)\; . \een

As is obvious the curvature of the two-sphere is not constant.
Only in the particular case $u_{D}= 0$ we obtain a constant
curvature.

The area of the horizon can be easily calculated and the result is

\ben {\cal A}_{\cal H}= 4\pi R_{{\cal H}}^2\left(1 - {Q^2\over
{\cal M}R_{{\cal H}}} \right) e^{-2(u_{*} + \upsilon_{1})} \;
.\een

The dual of the Maxwell tensor evaluated on the black hole horizon
is

\ben \star F\mid_{\,\cal H} = Q e^{-\upsilon_{2} - \upsilon_{1} }
\left(1 - {Q\over {\cal M}R_{\cal H} } \right)\sin(\theta)d\theta
\wedge d\phi  \; .\een

The charge of the distorted black hole is given by

\ben Q_{\cal H}= {1\over 4\pi} \oint_{\cal H} e^{-2\varphi}\star
F \,\,\, . \een

Performing integration on the horizon we obtain the explicit
expression

\ben Q_{\cal H}= Q e^{\upsilon_{2} - \upsilon_{1}} \; . \een

\section{Thermodynamics of distorted charged dilaton black holes}

In this section we discuss the zeroth and the first laws of
distorted charged dilaton black hole mechanics.

\subsection{Zeroth law}

The zeroth law states that the surface gravity is constant over
the horizon. The surface gravity is defined as

\ben \xi^{a}\nabla_{a}\xi^{b}= \kappa \, \xi^{b} \een

where $\xi$ is the horizon-generating Killing field. It should be
mentioned that  the above definition gives the surface gravity up
to a constant because there is a freedom to rescale $\xi$ by a
constant. Although, this rescaling does not influence the zeroth
law, for simplicity we assume from now on that our solutions have
been extended to be asymptotically flat. Then the freedom to
rescale the Killing vector can be fixed, requiring that the
Killing vector becomes time translation at infinity.

The explicit form of the surface gravity can be easily calculated
 and the result is

\ben \kappa = {1\over 4{\cal M}}e^{2(u_{*} + \upsilon_{1})}
\,\,\, .\een

The above result shows that the surface gravity is indeed constant
over the horizon and therefore the zeroth law is satisfied.

Moreover, as one may expect, the electric potential an the dilaton
turn out to be constant over the black hole horizon. Indeed, it is
not difficult to show  that

\ben \Phi_{\cal H}= {Q \over 2{\cal M}}e^{\upsilon_{1} -
\upsilon_{2}} \,\,\,\,\, , \,\,\,\,\,  e^{2\varphi_{\cal H}}=
e^{-2\upsilon_{2}} \left(1 - {Q^2 \over 2{\cal M}^2} \right). \een

\subsection{First law}

Variation of the black hole mass between neighboring equilibrium
states yields the first law of black hole physics. We want to
write down the first law for distorted charged dilaton black holes
regarded as  single systems acted on by external matter. The form
of the first law, of course, depends on how the black hole mass is
defined \cite{GH},\cite{FK}. We have three possible choices for
black hole mass in our case. The first natural choice is the ADM
mass ${\cal M}_{ADM}$ of space-time. The second choice is the mass
parameter ${\cal M}$ appearing in the black hole solution and the
third choice is the Komar mass of the horizon ${\cal M}_{\cal H}=
{1\over 4\pi}\kappa {\cal A}_{\cal H}$.

The ADM mass ${\cal M}_{ADM}$ contains a contribution from the
distorting external matter.  While we consider the distorted
charged dilaton black holes as single systems the ADM mass is not
appropriate for our purpose. The Komar mass ${\cal M}_{\cal H}$ is
just the gravitational mass of the black hole and does not contain
a contribution from the electromagnetic and dilatonic hair of the
black hole. Therefore, it is not appropriate, too. Only the mass
parameter ${\cal M}$ is appropriate in our case. The parameter
${\cal M}$ can be interpreted physically as "the mass of the black
hole alone as measured at infinity". This interpretation is in
agreement with the fact that ${\cal M}$ satisfies a Smarr formula:
${\cal M}= {\kappa\over 4\pi}{\cal A}_{\cal H} + \Phi_{\cal
H}Q_{\cal H}$.

Varying the mass ${\cal M}$ we obtain

\ben \la{FL} \delta {\cal M} = {1\over 8\pi} \kappa \delta {\cal
A}_{{\cal H }} +   \Phi_{{\cal H}}\delta Q_{{\cal H}} + {\cal
M}_{{\cal H}}\delta u_{*}  + {\cal M}\delta\upsilon_{1} + ({\cal
M}_{{\cal H }} - {\cal M})\delta \upsilon_{2} \een

This algebraic identity is the black hole first law for an
observer at infinity. The first two terms on the right-hand side
of (\ref{FL}) are the standard terms describing the change of the
mass due to the change in the area and the charge of the black
hole. There are, however, three new terms related to the
parameters $u_{*}$, $\upsilon_{1}$ and $\upsilon_{2}$.

The parameter $u_{*}$ comes from the Schwarzschild  solution and
thus it may be viewed as a quantity describing how the uncharged
and non-dilatonic external matter affects the black hole.
Moreover, $u_{*}$ couples to the pure gravitational energy ${\cal
M}_{\cal H}$ of the black hole and therefore the term ${\cal
M}_{\cal H}\delta u_{*}$ (as in \cite{GH}, \cite{FK}) can be
interpreted as the gravitational work done by uncharged and
non-dilatonic matter on the black hole.

The parameter $\upsilon_{1}$ can be considered as a quantity
describing the influence of the charged external matter on the
black hole. As can be seen from eqs. (\ref{GP}) - (\ref{EP}),
$\upsilon_{1}$ affects both the gravitational and electric
potentials and therefore the gravitational and electric parameters
on the horizon. That is why it sounds naturally that
$\upsilon_{1}$ couples to the total mass of the black hole which
contains contributions from gravitational, electromagnetic and
dilaton fields. Hence we can interpreted the term ${\cal M}\delta
\upsilon_{1}$ as the work done on the black hole by variations in
 the charged external matter.

The parameter $\upsilon_{2}$ can be considered  as a quantity
describing how the external dilatonic matter affects the black
hole. As can be seen  from eqs. (\ref{GP}) - (\ref{EP}), the
parameter $\upsilon_{2}$ affects the electric potential and
dilaton field but not the gravitational potential. That is why it
is reasonable that the parameter $\upsilon_{2}$ couples only to
the mass of the dilaton-electromagnetic hair of the black hole
but not to the gravitational part of mass. This is a consequence
of the well-known fact the black hole can not support pure scalar
hair. So, the term $({\cal M}_{\cal H} - {\cal M})\delta
\upsilon_{2}$ can be viewed as the work done on the black hole by
the external dilatonic matter.

The first law for distorted dilaton black holes obtained in the
present work can be considered as a natural generalization of the
first law for undistorted dilaton black holes found by Ashtekar
and Corichi in the isolated horizon context \cite{AC}. Here we
have considered the first law from the standard classical  point
of view. The general treatment of the first law in the isolated
horizon framework can be found in \cite{AFK} where some
particular examples as Einstein-Maxwell theory and EMD gravity
are considered. It should be noted, however, that no exact
distorted dilaton black hole solutions are presented in
\cite{AFK}.

\section{Conclusion}

In the present work we have constructed exact static, axisymmetric
solutions of the Einstein-Maxwell-dilaton gravity presenting
distorted charged dilaton black holes. The method for constructing
EMD solution is based on the global symmetry group of the
dimensionally reduced EMD equations. Making use of the group
action we can transform any given distorted black solution of the
pure Einstein equations into a static, axisymmetric  solution of
EMD equations. The new solution presents a distorted charged
dilaton black hole and depends on three arbitrary parameters. One
of the parameters comes directly from the seed distorted
Schwarzschild solution and describes how the uncharged and
non-dilatonic external matter affects the black hole. Two
remaining parameters are  group parameters. They can be
physically interpreted as quantities  describing  correspondingly
how the charged and dilatonic external matter affects the black
hole.

The zeroth and the first law of black hole thermodynamics have
been discussed, too.

It is worth noting that the method presented in this work can also
be applied  within the framework of Einstein-Maxwell gravity to
generate distorted charged black holes. Indeed,  after a
dimensional reduction the static, axisymmetric Einstein-Maxwell
equations have $SL(2,R)$ as a group of global symmetry. The
corresponding subgroup, preserving asymptotic flatness is again
$SO(1,1)$. Moreover, the method and the results of this work can
be generalized for EMD gravity with arbitrary dilaton coupling
parameter including as particular cases  the model considered here
and Einstein-Maxwell equations.

We would also like to discuss briefly  the following question. We
have generated the distorted charged dilaton black hole solutions
starting with the distorted Schwarzschild solutions. One may,
however, wonder whether the distorted charged dilaton black hole
solutions can be generated from the distorted black hole
solutions of Einstein-Maxwell gravity. We shall demonstrate that
this is possible without going into the details of calculation.

Let us return to the static, axisymmetric EMD equations
(\ref{SASEMD}) and to introduce the new potentials $U = u +
\varphi$, $\Psi= u - \varphi$, $\Lambda= \sqrt{2}\Phi$ and the
new metric function $H = 2h$. The static, axisymmetric EMD
equations can be rewritten in the form

\ben
\partial^2_{\rho}\Psi + {1\over \rho}\partial_{\rho}\Psi +
\partial^2_{z}\Psi = 0 \nonumber \\
\partial^2_{\rho}U + {1\over \rho}\partial_{\rho}U +
\partial^2_{z}U = e^{-2U}\left((\partial_{\rho}\Lambda)^2 + (\partial_{z}\Lambda)^2 \right)
\nonumber \\
\partial_{\rho}\left(\rho\partial_{\rho}\Lambda e^{-2u - 2\varphi}\right)
+ \partial_{z}\left(\rho\partial_{z}\Lambda e^{-2u -
2\varphi}\right)= 0 \\
 {1\over \rho} \partial_{\rho}H = (\partial_{\rho}U)^2 -
(\partial_{z}U)^2 + (\partial_{\rho}\Psi)^2 - (\partial_{z}\Psi)^2
- e^{-2U}\left((\partial_{\rho}\Lambda)^2 -
(\partial_{z}\Lambda)^2 \right) \nonumber  \\ {1\over
\rho}\partial_{z}H = 2\partial_{\rho}U \partial_{z}U   +
2\partial_{\rho}\Psi
\partial_{z}\Psi  - 2e^{-2U}\partial_{\rho}\Lambda \partial_{z}\Lambda
\nonumber \,\,\, .\een

It is not difficult to recognize that the above system is just the
static, axisymmetric   Einstein-Maxwell equations along with
minimally coupled scalar field $\Psi$. Let us denote by $U_{0}$
and $\Psi_{0}$ the potentials presenting the isolated charged
dilaton black hole. Note that $U_{0}$ is just the
Reissner-Nordstr\"om potential. Having once the potential $U=
U_{0} + U_{D}$ describing a distorted Reissner-Nordstr\"om black
hole we may construct immediately a distorted charged dilaton
black hole solution taking the pair of potentials $U$ and
$\Psi_{0}$. This solution, however, is not the most general one.
The most general charged dilaton black hole solution can be
obtained by adding to  $\Psi_{0}$ an arbitrary regular distortion
harmonic function $\Psi_{D}$.

In this paper we have consider only pure electrically
(magnetically) charged distorted dilaton black holes. Distorted
dilaton black holes with both electric and magnetic charges,
however, are not a trivial generalization of the above
considerations.  They form more complicated systems because they
can support additional axion hair.

\bigskip

\bigskip

\noindent{\Large\bf Acknowledgments}

\vskip 0.3cm

The author would like to thank P. Fiziev and V. Rizov for
stimulating discussions.


\bigskip


\begin{thebibliography}{31}

\bibitem{GH} R. Geroch, J. Hartle, J. Math. Phys. {\bf 23}, 680 (1981)

\bibitem {MS} L. Mysak, G. Szekeres, Can. J. Phys. {\bf 44}, 617
(1966)

\bibitem{P} P. Peters, J. Math. Phys. {\bf 20}, 1481 (1979)

\bibitem {X} B. Xanthopolous, Proc. Roy. Soc. Ser. {\bf A388}, 117 (1983)

\bibitem {FK} S. Fairhurst, B. Krishnan, Distorted black holes
with charge, Preprint gr-qc/0010088

\bibitem{Y1} S. Yazadjiev, Int. J. Mod. Phys. {\bf D8}, 635 (1999)

\bibitem{Y2} S. Yazadjiev, Exact static solutions in
Einstein-Maxwell-dilaton gravity with arbitrary dilaton coupling
parameter, gr-qc/0101078

\bibitem{AC} A. Ashtekar, A. Corichi, Class. Quantum Grav.{\bf 17},
1317 (2000)

\bibitem{AFK} A. Ashtekar, S. Fairhurst, B. Krishnan, Phys. Rev. {\bf
D62}, 104025-1 (2000)


\end{thebibliography}
\end{document}